# Elucidating the role of ribosomal A1493 in stabilization and rigid support for the codon-anticodon helix from molecular dynamics simulations

Shuhao Zhang[1], Zhen Wang[1], Jie Qiao[2], Wenbing Zhang∗

Department of Physics, Wuhan University, Wuhan, Hubei, People's Republic of China

Department of Science, Qingdao University of Technology, Qingdao, Shandong, China

## ABSTRACT

The ribosome ensures translational accuracy by monitoring codon-anticodon interactions at the A site decoding center. However, the mechanism by which conserved nucleotide A1493 contributes to this process remains controversial. To address this, we performed molecular dynamics simulations initiated from ribosomal recognition intermediate, revealing A1493's multiple roles in decoding. Through 2D umbrella sampling, we quantified the coupling between A1493 flipping and the stability of the first base pair in the codon-anticodon helix. Our results demonstrate that A1493 promotes mRNA-tRNA pairing through entropy-driven stabilization, amplifies stability differences between cognate and near-cognate tRNAs, and functions as a "wedge" to provide rigid support for the codon-anticodon helix. Further analysis identified steric complementarity as essential for proper A1493 flipping, with the tRNA 37th position playing a key structural role. Additionally, we explored how Ψ modification and translocation influence codon-anticodon stability. Together, these findings refine the current mechanistic framework of ribosomal decoding.

## INTRODUCTION

In protein synthesis, the elongation factor (EF-Tu) delivers aminoacyl-tRNA (aa-tRNA) to the ribosomal decoding center as an aa-tRNA•EF-Tu•GTP ternary complex, enabling mRNA codon recognition on the 30S subunit (1–3). The decoding center does not lock tRNA at the initial recognition but dynamically monitors codon-anticodon interactions before and after GTP hydrolysis (4–6). Mismatched tRNAs are rejected and dissociate (7), whereas cognate tRNA binding accelerates EF-Tu GTPase hydrolysis and accommodation (4, 8–10). This process triggers 30S "domain closure"—a head-to-shoulder rotation tightening the acceptor site (2, 9). Concurrently, aa-tRNA enters the 50S large subunit's A-site for peptide bond formation, followed by translocation mediated by elongation factor (EF-G) to prolong the synthesis of peptide chain (11–15).

   In the above process, precise discrimination between cognate and near/non-cognate tRNAs by the decoding center is fundamental for translation fidelity. Misselection events induce amino acid substitutions that may disrupt protein functionality. Extensive studies have demonstrated that several conserved ribosomal nucleotides—including G530, A1492 and A1493 of 16S rRNA, and A1913 of 23S rRNA—are critical for this process (2, 16–19). Structural analyses reveal that A1493 forms A-minor interactions with the first base pair of the codon-anticodon helix, while A1492 and G530 engage with the second and third base pairs, respectively. However, the mechanism by which these interactions enhance decoding accuracy remains contentious. One model proposes that A1492/A1493 flip out from helix 44 to dynamically monitor Watson-Crick geometry at the first two base pairs, stabilizing cognate tRNA through minor-groove interactions and inducing domain closure (10, 20–23). In contrast, near/non-cognate tRNA binding induces non-Watson-Crick conformations that disrupt these groove interactions, which consequently fail to induce A1492/3 flipping and subsequent GTPase activation.

∗ *Corresponding author: Wenbing Zhang (E-mail: wbzhang@whu.edu.cn)*

However, this model was challenged by structural evidence showing that near-cognate tRNA-ribosome complexes adopt an identical decoding center structure to that of cognate complexes. In these near-cognate complexes, mismatches at the first two positions of the codon-anticodon helix maintain Watson-Crick base pairing (e.g., U-G, G-U) through keto-enol tautomerization (24–26); conserved residues G530, A1492, and A1493 interact with both cognate and near-cognate helices in the same manner; and the 30S subunit undergoes identical domain closure. These observations indicate that A1492/3 flipping lacks tRNA specificity and, together with codon-anticodon helix, forms a rigid decoding center, thereby enforcing Watson-Crick geometry at the first two base-pair positions. While, some research through molecular simulation have revealed flipping specificity and hydrogen bond variations between cognate and near-cognate systems (27). Chemical modifications that disrupt minor-groove hydrogen bonds have been shown to minimally affect tRNA discrimination, highlighting steric complementarity rather than hydrogen bonds as the key determinant (28, 29).

In addition, several experiments have revealed that various RNA modifications differentially regulate ribosomal translation processes (30). For example, m6A and m5C modifications increase translation efficiency by recruiting reading proteins (31, 32). However, certain RNA modifications such as pseudouridine ($\Psi$) and N1-methylpseudouridine (m1$\Psi$) in mRNA vaccines can significantly influence translation despite lacking corresponding reader proteins. In addition to its contribution to the stability of mRNA, single U-to-$\Psi$ substitutions in codon increase amino acid substitution and reduce GTP hydrolysis rates, while $\Psi$ incorporation at stop codons induces readthrough events (33). Although m1$\Psi$ modification shows no effect on codon misreading, it modulates protein synthesis rates and significantly enhances +1 ribosomal frameshifting (34). A comprehensive understanding of these modification mediated translation mechanisms is crucial to reduce the possible mistranslation in mRNA vaccines. To elucidate modification-induced amino acid substitution events, it is necessary to compare the structure of the near-cognate tRNA-ribosome complexes in the pre-dissociation state with those that evade ribosomal proofreading after modification.

However, the experimentally resolved crystal structure of the near-cognate tRNA-ribosome complex was obtained under conditions in which only the near-cognate tRNA was added to the solution, without the cognate tRNA (24–26). These resolved structures represent rare proofreading evade events under cellular conditions, correlating with the low intrinsic amino acid substitution rate ($10^{-4} \sim 10^{-3}$) observed in vivo (35). It remains unclear what structural features of the vast majority of near-cognate tRNA-ribosome complexes that are rejected by ribosomal proofreading. Several questions persist: Do these rejected near-cognate tRNAs form Watson-Crick base pairing with mRNA codons, only to dissociate subsequently due to unfavorable isomerization energy penalties? Or does rejection result from failed canonical base pairing? Do these conserved nucleotides in the decoding center maintain similar conformations during rejection events? Furthermore, what is the individual contribution of each conserved nucleotide to the stability of the codon-anticodon helix? What conformations do mismatch base pairs adopt in near-cognate systems when either the codon or anticodon is chemically modified? These questions may require computational simulations for resolution, as experimentally capturing transient structures of rejected near-/non-cognate tRNAs during ribosomal discrimination remains challenging.

In this study, we employed molecular dynamics simulations to investigate the energetics of microscopic events during tRNA recognition. Our simulation system was constructed based on the pre-accommodation intermediate structure (PDB: 5UYL) reported by Loveland et al. (18), capturing the ribosome complex post EF-Tu•GTP•aminoacyl-tRNA binding but preceding 30S subunit closure. Within the decoding center, the anticodon base pairs with the codon, with A1492 resides inside helix 44 and stacks with A1913 from helix 69 of 23S rRNA, while A1493 adopts a flipped-out conformation interacting with the minor groove of the first base pair of codon-anticodon helix.

We focused on A1493 and the first-position base pair of the codon-anticodon helix that interacts with A1493. Using one-dimensional and two-dimensional umbrella sampling, we investigated (1) the stability changes of the first-position base pair, (2) the flipping angle changes of A1493, and (3) their coupling relationship, when codon with U at first position pairs with either cognate or three near-cognate tRNAs. Our results demonstrate that A1493 enhances mRNA-tRNA pairing through entropic stabilization, amplifying stability differences between cognate and near-cognate complexes. And A1493 functions as a "wedge", providing rigid support for the backbone of the codon-anticodon helix. We confirmed the importance of steric complementarity for proper A1493 flipping and elucidated the structural role of the tRNA's $37^{th}$ nucleotide during this progress. These results indicate a complementary relationship between the two prevailing models. Furthermore, we found that U to $\Psi$ modification stabilizes backbone distortion at the P/A kink via water-mediated bridging, thereby enhancing codon-anticodon pairing. This provides a structural basis for the experimentally observed $\Psi$-induced amino acid substitution. Finally, we explored stability changes in the codon-anticodon helix during ribosomal translocation, revealing both transient destabilization during translocation and the stabilizing contribution of U1498 of helix 44 to P site stability.

## MATERIALS AND METHODS

### System setup

To simulate the decoding center of the ribosome's A site, we extracted atoms within a 20Å linear distance from both A1493 and the first-position base pair in the 70S crystal structure (PDB: 5UYL), as shown in Figure 1. The rRNA is colored blue, the A-site tRNA yellow, the mRNA red, and the P-site tRNA orange. Given the nearly identical structural conformations adopted by diverse tRNAs upon A-site binding, we mutated the A-site tRNA sequences to four variants: Phe-tRNA (anticodon 5'-GAA-3'), Ile-tRNA (5'-GAU-3'), Val-tRNA (5'-GAC-3'), and Leu-tRNA (5'-GAG-3'). Full tRNA sequences were obtained from the *E. coli* Modomics database (36; see SI). The mRNA sequences were 5'-AUG[UUU]A-3' and 5'-AUG[ΨUU]A-3', with bracketed codons. This yielded eight A-site ribosomal systems with the first-position base pair (U/Ψ)-(A/U/C/G) and an invariant U-A pair at the second position (mRNA base listed first). To assess ribosomal effects, control models of mRNA-tRNA duplexes lacking ribosomes were constructed, solvated identically, with restraints applied to the terminal two bases of mRNA and tRNA to maintain the structure in ribosomes. Four ribosome-free systems were generated, featuring first-position (U/Ψ)-(A/U) base pairs. Additional A-site models were built using crystal structures 4V9I and 4JV5(37), capturing A1493 in anti/syn conformations with Ψ-A first base pairs. The ribosomal P-site model in the post-translocation state was constructed using crystal structure 7SSW(15), retaining atoms within a 20 Å radius of the first base pair of P-site codon.

## MD simulations

Molecular simulations were conducted using the GROMACS-2022. The protein were parameterized with the ff14SB force field (38), while standard RNA nucleotides were modeled using the OL3 force field (39, 40). Pseudouridine modifications were described with the latest force field proposed by Dutta et al.(41), and other RNA modifications in tRNA were described with the Modrna08 force field (42). Systems were charge-neutralized with Na+ ions and adjusted to 150 mM ionic concentration using NaCl. The ribosome A-site model was solvated in a dodecahedral box of TIP3P water molecules, with the box boundaries positioned at a minimum distance of 10 Å from the model. Electrostatic interactions were calculated using the Particle Mesh Ewald (PME) method with a cutoff distance of 12 Å. Van der Waals (Lennard-Jones) interactions were truncated at 12 Å. All bonds involving hydrogen atoms were constrained using the LINCS algorithm. To preserve the structural integrity of the ribosome fragment, outer heavy atoms in the solute model located more than 17 Å from A1493 and the first base pair were restrained to their initial positions, while the inner region remained flexible. The system underwent energy minimization, followed by 2 ns of NVT ensemble heating to 310 K using the Berendsen thermostat, 5 ns of NPT ensemble density equilibration with the Berendsen barostat, and an additional 20 ns of NVT equilibration to obtain the initial structure for subsequent simulations. Statistical data for all unbiased simulations were collected over 500 ns in the NVT ensemble.

## US simulations

The free energy profiles were computed using umbrella sampling (US), including the one-dimensional free energy surface for the opening distance of the first base pair in the codon-anticodon helix, the one-dimensional free energy surface for the flipping angle of A1493, and the two-dimensional free energy surface combining both variables. The distance of the first base pair was restrained with a force constant of 2000 kJ/mol/nm², incrementally increasing from 5 Å to 12 Å in 0.5 Å steps. The flipping angle of A1493 was restrained with a force constant of 500 kJ/mol/rad², decreasing from 170° to 60° in 10° intervals. To prevent complete unwinding of the codon-anticodon duplex during the gradual opening of the first base pair, a wall restraint of 500 kJ/mol/nm² was applied when the distance between bases in the adjacent second base pair exceeded 0.1 Å beyond their equilibrium distance. Each window in the two-dimensional umbrella sampling was simulated for 22 ns, with the final 20 ns analyzed using the Weighted Histogram Analysis Method (WHAM) (43) to compute the unbiased free energy surface. For one-dimensional umbrella sampling, each window was simulated for 42 ns, with the last 40 ns used for free energy calculations. Statistical errors were estimated via block analysis, dividing each window's trajectory into four segments and calculating the standard error of the free energy across these segments.

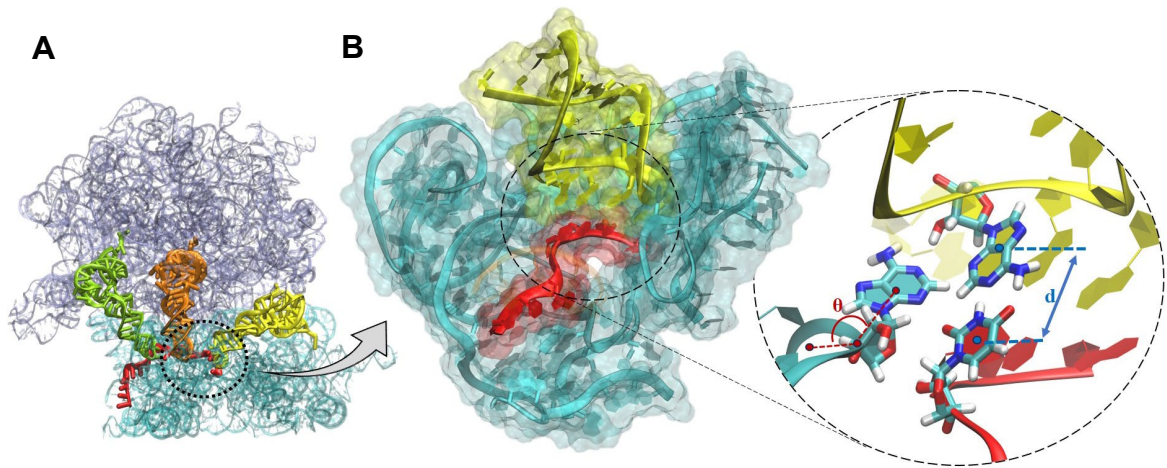

**Figure 1**. Preparation of the A-site models. (A) Visualization of the A-site model (right) and its position and size in the 70S ribosome (left). (B) Atoms within 20 Å linear distance from both the first base pair of the A-site codon-anticodon helix and nucleotide A1493 (demarcated by black circle in panel A). Color scheme: rRNA (blue), A-site tRNA (yellow), mRNA (red), and P-site tRNA (orange). The inset panel shows a close view of the first base pair-A1493 interaction interface, with annotated reaction coordinates for umbrella sampling simulations.

## RESULTS

**The decoding center structure of near-cognate complexes**

We constructed four tRNA-ribosome complexes, Phe-tRNA, Ile-tRNA, Val-tRNA and Leu-tRNA, in which the first base pair of the codon-anticodon helix is U-A, U-U, U-C, and U-G, respectively. Initial investigations focused on whether first-position mismatches alter helical geometry or induce groove width variations. As shown in Table 1, the groove width of these four pairs does not change much, and the helix remains in the A-form. Figure 2 shows U-U and U-C mismatches each form a single hydrogen bond, whereas the U-G pair form two hydrogen bonds through slight major-groove flip of U4 (the first codon base), adopting a wobble base-pair structure.

Experimental studies have identified two U-G conformations in near-cognate systems, the first is the classical wobble pairing supporting the dynamic decoding model, and the second is a Watson-Crick configuration underlying the rigid model (2, 25). The rigid model posits that P/A kink-induced torsional constraints prevent U4 flipping. However, our equilibrium simulations show that U4 undergoes slight major-groove flipping despite the P/A kink, enabling wobble pairing. This suggests that the torsional constraints imposed by the P/A kink cannot fully restrict U4's slight conformation adjustments. Therefore, we conclude that the wobble base-pairing constitutes a pre-dissociation state rejected by ribosomal proofreading.

Notably, our simulated U-U configuration differs from the experimental structure in PDB:5EL4. In this experimental structure, the U-U pair maintains a Watson-Crick geometry but with elongated hydrogen bonds (3.4 Å and 3.7 Å), leading to the conclusion that keto-enol tautomerization did not occur (24). Given that the adjacent base pair U5-A35 in our simulated system differs from A5-U35 in 5EL4, we reconstructed the simulation system based on 5EL4. Under the assumption of no keto-enol tautomerization, the Watson-Crick U-U configuration destabilized rapidly (initiating at the energy minimization stage), with the mRNA U4 undergoing minor groove flip to alleviate electrostatic repulsion between the 3′-NH and 4′-oxygen (Figure S1). This indicates that the 5EL4 system must have undergone tautomerization and form two weak hydrogen bonds. Detailed hydrogen-bonding networks for all four systems are analyzed in the Supplementary Information (Figure S2).

**Table 1.** Structural parameters of first base pair in 500 ns equilibrium simulations for different systems

| Base Pair | Backbone Spacing [Å] | Base Spacing [Å] | A1493 Angle (°) |
|---|---|---|---|
| U-A (Phe) | 11.0 ± 0.2 | 5.9 ± 0.1 | 138.1 ± 4.8 |
| U-U (Ile) | 11.1 ± 0.5 | 6.1 ± 0.3 | 132.5 ± 8.5 |
| U-C (Val) | 11.3 ± 0.5 | 6.2 ± 0.5 | 130.3 ± 4.3 |
| U-G (Leu) | 10.7 ± 0.2 | 5.9 ± 0.2 | 128.5 ± 4.6 |

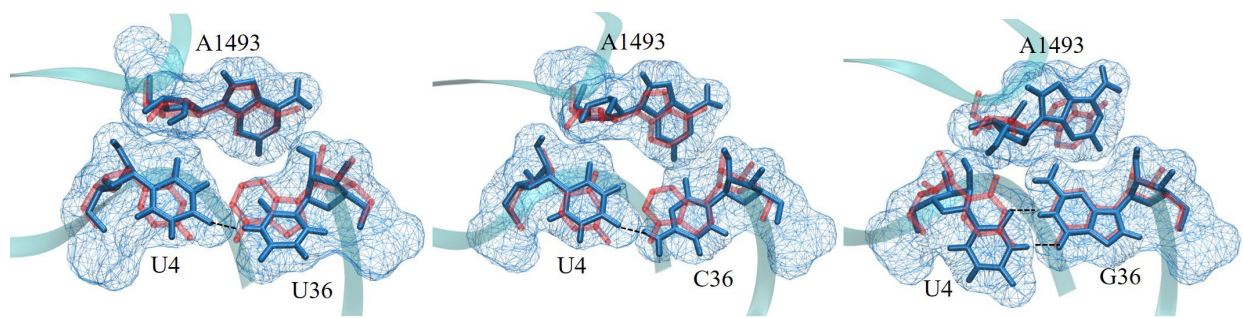

**Figure 2**. Structural comparison of decoding centers between the near-cognate system (blue) and the cognate U-A pair (red). Hydrogen bonds formed by base pairing are indicated with black dashed lines.

**Mechanism of A1493 in stabilizing the codon-anticodon helix**

While ribosome is known to stabilize cognate tRNA, the quantitative aspects of this stabilization remain poorly characterized. To address this, we performed one-dimensional umbrella sampling calculations to quantify the base pair opening free energy of the mRNA-tRNA duplex at the first codon position, both within the ribosome and in its absence (Figure 3A). The free energy profile for U-A pair opening within the ribosome (blue curve) exhibits significantly greater stability compared to the ribosome-free system (purple dashed curve).

When the umbrella bias gradually pulls U4 and A36 apart, A36 of tRNA is almost motionless, and only U4 of mRNA moves away from A36. This rule is consistent with the presence or absence of ribosomes. This shows that the first codon of mRNA is unstable relative to the anticodon of tRNA. Although the anticodon region of tRNA is a loop as shown in Figure

S3, the structure is very compact, with adjacent bases stacked on each other; and a U-turn occurs at U33, which is stabilized by the hydrogen bond between the N3 atom of the U33 base and the phosphate group of A36. This hydrogen bond also limits the backbone flexibility of A36.

The A-site codon and the P-site codon of mRNA form two A-form small helices with different axes, creating a pronounced kink at their junction—the P/A kink (Figure 3B). This structural discontinuity disrupts the stacking stabilization of U4 from its 5'-adjacent base. Simulations of isolated mRNA (lacking A-site tRNA pairing) with constrained P-site codon and the last two A-site bases (mimicking ribosomal mRNA conformation) revealed that torsional stress accumulated at the P/A kink destabilizes the U4-U5 base stacking. As a result, U4 rapidly flips into solvent during initial simulation stages (transparent structure in Figure 3B), demonstrating that the P/A kink directly destabilizes the first codon position. This destabilization is partially counteracted by cross-strand stacking between U4 and the tRNA's 37th base. The 37th position of tRNA is often hypermodified to stable the weak A-U base pair at the first position of codon, which is the current general understanding of the role of the 37th position of tRNA. In Phe-tRNA, for instance, the native A37 is hypermodified to SPA, where its sulfur atom forms π-S interactions with U4 base. Energy calculations from molecular force fields revealed that unmodified A37 interacts with U4 at -3.1 kcal/mol, whereas SPA modification strengthens this interaction to -5.5 kcal/mol, surpassing the U5-U4 stacking energy (-3.7 kcal/mol). However, this cross-strand stabilization still cannot compensate for the destabilization induced by the P/A kink. Combined with the easier opening nature of the 5' end of RNA (43) and the relatively weak stacking propensity of uracil, U4 is still the source of instability of this base pair.

The opening pathways of U4 in ribosome (blue scatter) and ribosome-free (orange scatter) systems are depicted in Figure 3C. The horizontal axis is the U4 flipping angle (defined by the dihedral angle formed by the mass-center of U5 base, U5 sugar, U4 sugar, and U4 base), while the vertical axis is the U4 opening distance. At equilibrium, the U4 flip angle is about 20°, with angle increases correspond to base flipping toward the minor groove and decreases toward the major groove. It can be clearly seen from the figure that the two systems follow two different opening pathways. In ribosome, the backbone distortion of the P/A kink forces U4 to flip toward the minor groove, but this will cause a steric conflict with A1493, so U4 can only flip into the solution after moving a certain distance along the U-A base pair plane (the flip angle of the blue pathway does not change at the beginning). And in this process, the 2'-OH of the U4 sugar has been forming a hydrogen bond with the phosphate group of A1493, further restricting the conformation of U4 in the open state. In contrast, the ribosome-free system shows immediate flipping into solvent from minor groove once U4 opening distance reaches ~6.7 Å, with disruption of U5 stacking and A36 hydrogen bonding, thereby achieving open state.

Figure 3D displays structural superpositions of both systems at a 7.5 Å opening distance. In the ribosome-free system (orange structure), U4 flips into solvent, corresponding to the open state past the transition state on the free energy surface (purple dashed line in Figure 3A). Notably, this open conformation sterically conflict with the sugar of A1493 and the phosphate group of G1494 after superposition (depicted as dark blue molecular surfaces). Conversely, U4 remains constrained in the ribosome decoding center (blue structure), keeping the orientation of the base plane in the closed state unchanged, which is reflected in the transition state corresponding to the free energy maximum (blue solid line in Figure 3A). In summary, ribosome A1493 reduces the entropy increase of the opening process by limiting the conformations numbers of open state of the first base of the codon, making U4 more inclined to maintain closed state, thus stabilizing the interaction between mRNA and tRNA, which is an entropy stabilization mechanism.

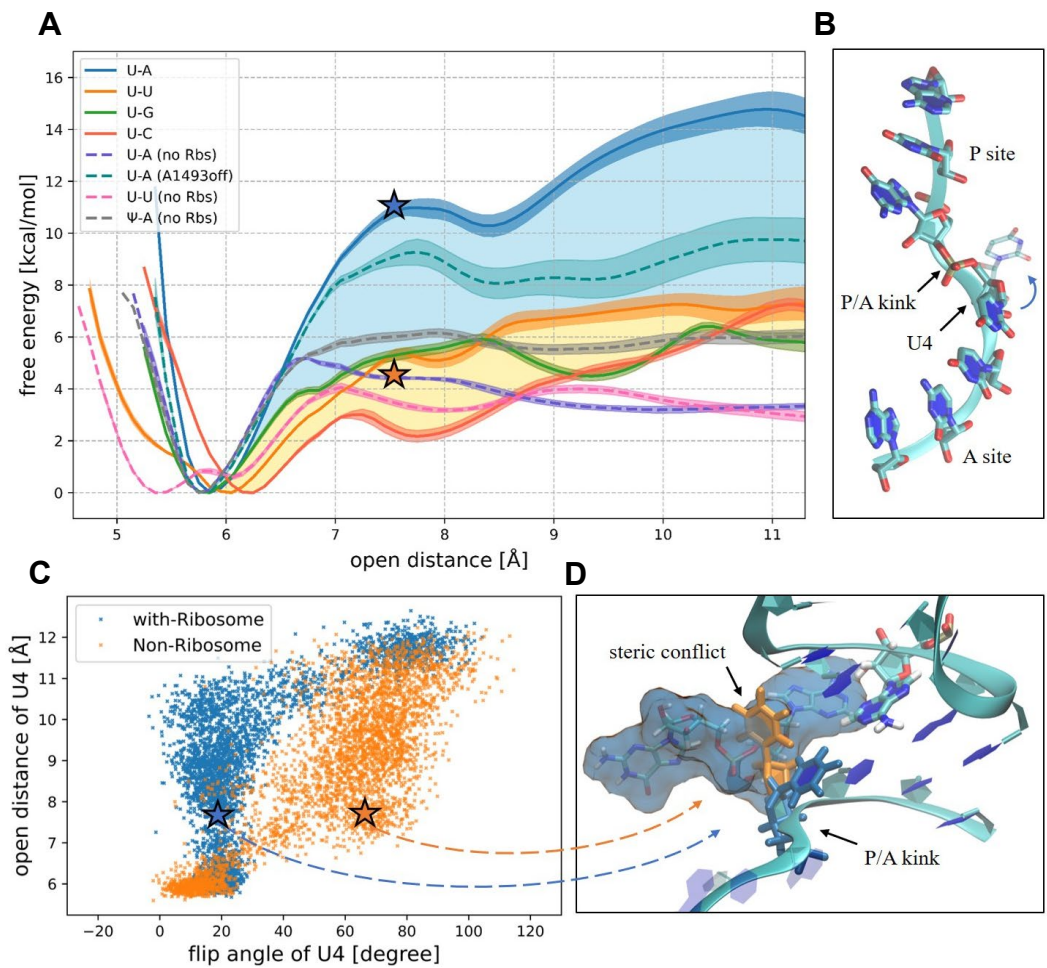

**Figure 3**. Free energy and structure of the first base pair. (A) Free energy landscapes of first base pair opening for different systems. Solid lines: U-A/U/C/G pairs in ribosome-bound systems (blue/orange/green/red). Dashed lines: Ribosome-free U-A (purple), U-U (pink), Ψ-A (gray), and A1493-off U-A (cyan). Yellow/blue zone: free energy distributions for near-cognate systems and the discrimination gap between cognate system. Errors are shown as color strips. (B) mRNA single-strand mimic with constrained termini (3'/5') and flipped-out U4 (stransparent). (C) Scatter plot overlay of all trajectory windows for the first base pair in ribosome-bound (blue) and ribosome-free (orange) systems, mapped onto U4 flip angle versus U4 open distance. The two systems exhibit distinct opening pathways. (D) Structural comparison at 7.5 Å opening distance (corresponding to the star marked in panels A and C). The molecular surfaces of A1493 and G1494 are shown in dark blue, which produce steric conflicts with the flipped U4 in ribosome-free system.

**The stability contributed by A1493 depends on its flip angle**

During the above process of U4 opening, A1493 maintains interact with the minor groove of the codon-anticodon helix. We next investigate the contribution of different flip angles of A1493 to the stability of the first base pair. The flip angle of A1493, which characterize the extrahelical flipping from helix 44, is a dihedral angle composed of the following four center-of-mass: G1494 base pair and G1491 base pair; G1494 phosphate group; A1493 phosphate group; and A1493 base. Since this involves both the change in the flip angle of A1493 and the distance of the first base pair, we employed two-dimensional umbrella sampling to compute their coupling, as shown in Figure 4A. The two-dimensional free energy map reveals a concentrated distribution of A1493 flip angle and first base pair distance, with only one minimum.

When the flip angle of A1493 exceeds 120°, A1493 forms A-minor interactions with the codon-anticodon helix. As the flip angle decreases, A1493 gradually moves away from the codon-anticodon helix, accompanied by the disruption of A-minor interactions. At angles below 80°, the stacking between A1913 and A1492 is disrupted, and A1493 progressively returns to the interior of helix 44. Since the system is modeled based on the recognition intermediate (where A1913 is inserted into helix 44 and stacks with A1492), the flip angle of A1493 is restricted to greater than 60°. We categorize A1493 into two states: the "on" state, where the flip angle is greater than 120° (fully extrahelical flip), and the "off" state, where it is less than 100° (partially extrahelical flip). We obtain the one-dimensional free energy profiles for U4 opening when A1493 is in each of these states by integrating the two-dimensional free energy along CV2:

$$F_{on}(CV1) = -k_B T ln(\int_{CV2 \in on} e^{-F(CV1,CV2)/(k_B T)} dCV2) \quad (1)$$

$$F_{off}(CV1) = -k_B T ln(\int_{CV2 \in off} e^{-F(CV1,CV2)/(k_B T)} dCV2) \quad (2)$$

where CV1 corresponds to the distance between the first base pair, while CV2 denotes the flip angle of A1493. When A1493 is in the "on" state, the free energy profile for U4 opening closely aligns with the overall one-dimensional U4 opening free energy (blue solid line in Figure 3A). This alignment is expected, as the free energy minimum for A1493 occurs in the "on" state, indicating that the total U4 opening free energy is almost entirely contributed by A1493 in this state. In contrast, when A1493 is in the "off" state, the U4 opening free energy corresponds to the cyan dashed line in Figure 3A, exhibiting a lower free energy barrier compared to the "on" state, which suggests easier dissociation.

The structural transition of A1493 from the "on" state to the "off" state is depicted in Figure 4B, where the transparent structure represents the "on" state and the solid structure represents the "off" state. As A1493 flips back into the interior of helix 44, the spatial position of its sugar and backbone undergoes a slight shift, gradually moving away from U4. This movement allows U4 in the open state to have more accessible conformations. Consequently, compared to when A1493 is in the "on" state, U4 shows a higher tendency to adopt the open state, resulting in a corresponding decrease in the stability of the mRNA-tRNA interaction. This observation highlights the biological significance of the correct flip angle of A1493: its accurate flipping maximally stabilizes the codon-anticodon helix.

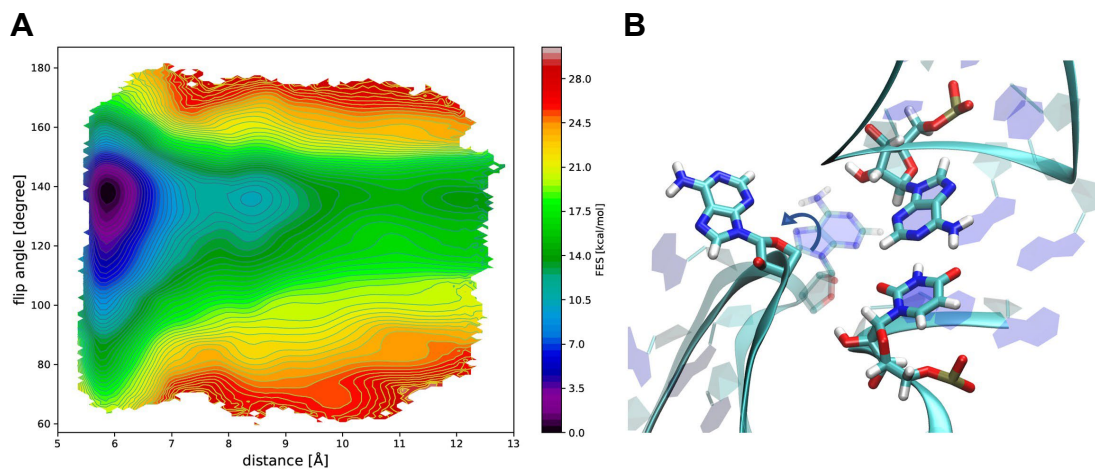

**Figure 4**. (A) 2D free energy map of U-A base pair distance and A1493 flip angle. (B) Structural changes from the on state (shown in transparent structure) to the off state of A1493.

**The rigid support of A1493 for the codon-anticodon helix**

Analysis of cognate tRNA systems in the previous section revealed a coupling between the flip angle of A1493 and the distance of the first base pair. Here, we investigate whether this coupling changes in near-cognate tRNA system.

Figure 5A presents the 2D free energy map of the U-U base pair opening distance and the A1493 flip angle in the near-cognate Ile-tRNA system. Compared to the cognate system, the free energy distribution is more diffuse, yet the free energy minimum remains nearly unchanged (see Table 1). The equilibrium distance of the U-U pair is about 6.1 Å, and only one hydrogen bond is formed. However, when the A1493 flipping angle decreases below 120° and gradually moves away from the minor groove of the codon-anticodon helix, the distance between U-U is also gradually decrease, as indicated by the red dashed trajectory. The free energy curves for U4 opening when A1493 is in the "on" state and "off" state are depicted by the orange and blue lines in Figure 5C, respectively. Notably, the free energy minimum for the U4 opening distance shifts. As A1493 transitions from the "on" state to the "off" state, the U-U backbone distance decreases from 11.1 Å to 9.1 Å (defined by the distance of atomic C1'), and the base pair distance reduces from 6.1 Å to 5.4 Å. Additionally, both bases exhibit a slight flip toward the major groove, favoring the formation of two hydrogen bonds (see inset in Figure 5C). This suggests that the U-U pair tends to form a stronger interaction (dual hydrogen bonds); however, due to the constraint imposed by A1493, the mRNA and tRNA backbones cannot approach too closely, maintaining a relatively larger distance and restricting the interaction to a single hydrogen bond. Therefore, A1493 acts as a "wedge", providing support between the mRNA and tRNA backbones.

For comparison, the U-U opening free energy curve in the absence of the ribosome, shown as the pink dashed line in Figure 3A, indicates that without A1493's support, the U-U base pair distance shortens to approximately 5.4 Å—similar to the U-U structure observed in the ribosome when A1493 is in the "off" state. Our findings demonstrate that A1493 effectively restricts the tRNA backbone, preventing excessive proximity to the mRNA, thus providing a rigid structure for the decoding center. This supports the rigid decoding center hypothesis proposed in the second model.

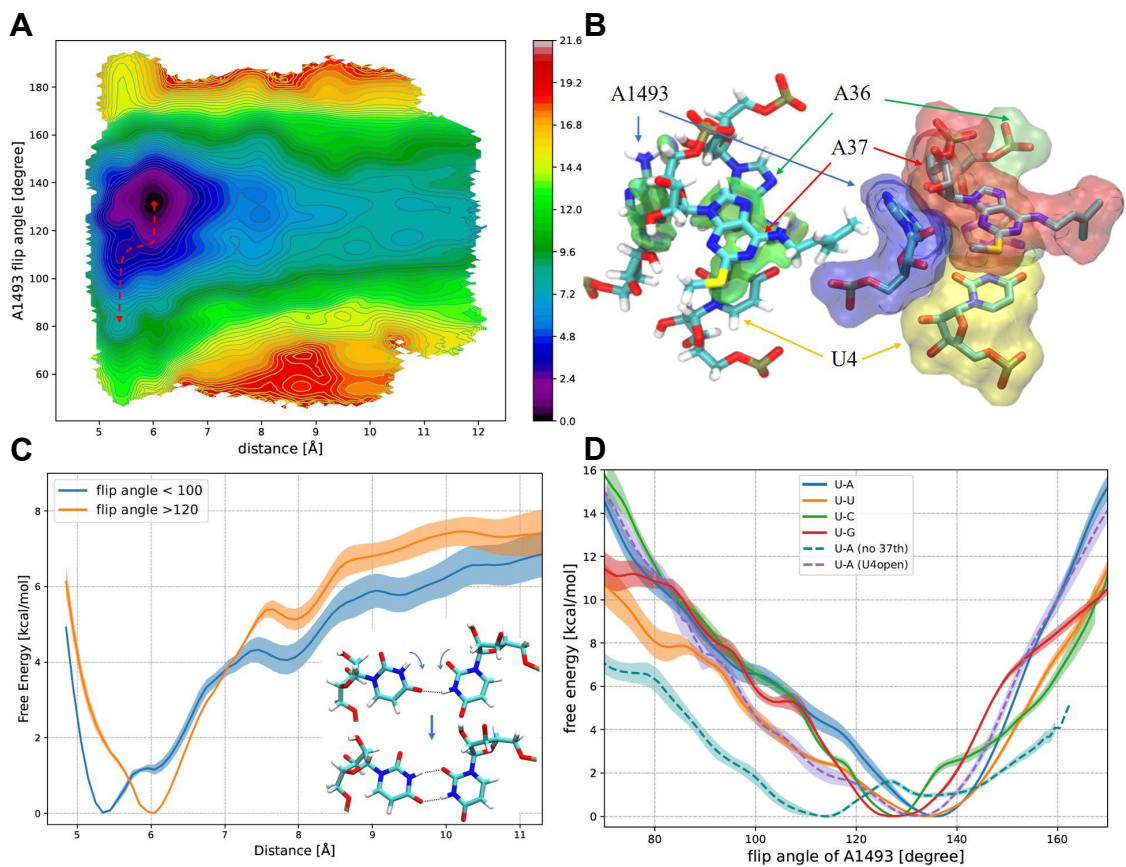

**Figure 5.** (A) 2D free energy map of U-U base pair distance and A1493 flip angle. The red dashed line shows the trajectory of the decrease in U-U distance as the flip angle of A1493 decreases. (B) Left: Visualization of weak interactions around A1493 calculated by aIGM, with the green surface representing van der Waals interactions. Right: Molecular surfaces of A1493, the first base pair, and the tRNA nucleotide at position 37, highlighting spatial complementarity within the decoding center. (B) Free energy curves for U4 opening in the A1493 "on" (orange) and "off" (blue) states, showing a shift in the free energy minimum. The inset depicts the structural transition of the U-U base pair from a single hydrogen bond (on state) to a double hydrogen bond configuration (off state). (D) Free energy curves for A1493 flipping in different systems. Blue, orange, green, and red solid curves correspond to systems with U-A, U-C, U-G, and U-U base pairs, respectively. The purple dashed curve represents the A1493 flipping free energy when the U-A base pair is in an open state, while the cyan dashed curve shows the profile in the absence of the tRNA nucleotide at position 37. Errors are shown as color strips.

**Steric complementarity and the structural role of the 37th site of tRNA**

To elucidate how A1493 functions as a "wedge", we employed the aIGM method and Multiwfn software (44–47) to visualize weak interactions surrounding A1493 in the cognate system (Figure 5B, left panel).

The analysis reveals a green surface of van der Waals (vdW) interaction formed between the base of A1493 and the sugar of the tRNA nucleotide at position 37, which structurally supports the tRNA backbone. These vdW interactions is sensitive to the adjacent spatial structure (because the attractive and repulsive terms scaling as $r^{-6}$ and $r^{-12}$, respectively). As shown in the molecular surface representation (Figure 5B, right panel), the first base pair and 37th site of tRNA collectively form a cavity that accommodates A1493. Although individual vdW interactions are weaker than hydrogen bond interaction, their cumulative effect—arising from the extended surface of steric complementarity—exceeds the strength of hydrogen bonding in the A-minor groove interaction. Notably, experimental studies have demonstrated that disrupting hydrogen bond in this region only have minimal impact on ribosomal decoding (28, 29), suggesting a prior overestimation of the functional role of A-minor hydrogen bonding. Instead, the "Lego-like" steric complementarity plays a dominant role.

We next constructed a control model by removing all nucleotides from the cognate tRNA except the anticodon, thereby eliminating interactions between A1493 and the 37th tRNA nucleotide (Figure 6A). The resulting free energy curve, is shown as cyan dashed curve in Figure 5D, revealed a significant shift in the free energy minimum for A1493 flipping: the equilibrium flipping angle shifted from 138° (intact anticodon loop system, blue solid curve) to approximately 117°, corresponding to a ~20° angular displacement. This energy minimum shift indicates that the absence of the 37th nucleotide induces a conformational rearrangement of A1493 in its stabilized equilibrium state. Structural comparisons of the two systems at 117° (Figure 6B, C) reveal that the absence of the tRNA 37th nucleotide induces rotation of the A1493 base plane (glycosidic angle changed from ~300° to ~180°) during its incomplete extrahelical flip. In contrast, when the tRNA 37th nucleotide is present, vdW interactions constrain A1493 without base plane rotation, enabling it to maintain steric complementarity with the tRNA and mRNA, thereby maximizing vdW stabilization. This reveals the role of the tRNA 37th nucleotide in stabilizing the A1493 conformation and confirms that steric complementarity is more critical than hydrogen bonding. When the spatial structure is destroyed, hydrogen bonding alone is not enough to recover the correct flip angle of A1493.

Additionally, when the U4 base pair adopts an open state (disrupting the Watson-Crick geometry), the free energy curve of A1493 flipping (purple dashed curve in Figure 5D) is close with that of the closed U4 state (blue solid curve), demonstrating that proper positioning of A1493 is maintained as long as one base in the first-position base pair retains its native conformation.

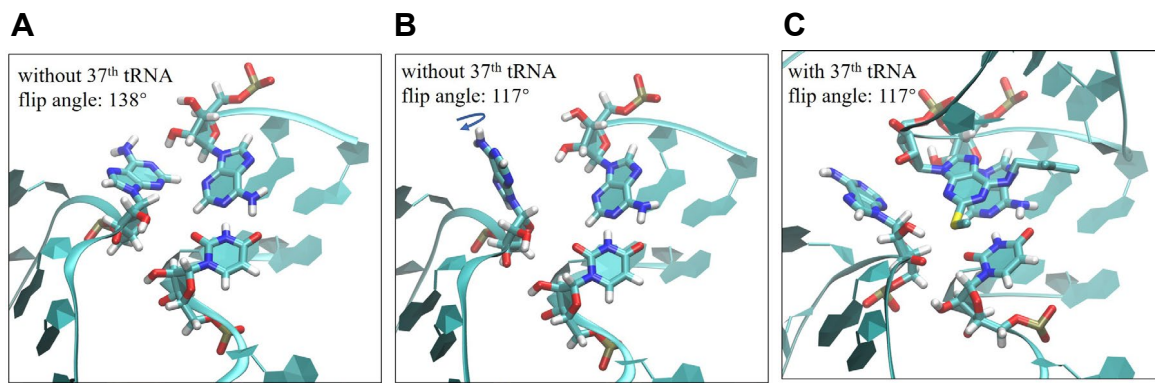

**Figure 6.** (A) Structure of the A1493 "on" state (flipping angle = 138°) when the tRNA retains only the anticodon nucleotides. (B) New stabilized conformation of A1493 (flipping angle = 117°) in the absence of the tRNA 37th nucleotide (anticodon-only system) with rotation of base plane. (C) Unstable conformation of A1493 (flipping angle = 117°) in the intact anticodon loop system.

**Ribosomal discrimination between cognate and near-cognate trna**

The opening free energy curves of the first base pair in three near-cognate systems are distributed in the relatively concentrated yellow zone in Figure 3A, where the orange, red, and green curves represent U-U, U-C, and U-G base pairs, respectively. There is a significant stability difference between them and the free energy curve of the cognate system U-A, represented by the light blue zone (about 6 kcal/mol), which is sufficient for the ribosome to distinguish.

However, the free energy difference between U-A and U-U base pairs in the absence of the ribosome is very small (about 1 kcal/mol, as shown by the purple and pink dashed lines in Figure 3A). This indicates that the ribosome can amplify the stability difference between the cognate system and the near-cognate system to effectively distinguish them. This amplification mechanism achieved through the desolvation of the decoding center by the ribosome (2, 48), the stabilization of the hyper-modified nucleotide at the tRNA37 position (such as the π-S interaction of Phe-tRNA), and the stabilization of A1493. Therefore, it does not support the previous view that the ribosome has the same stabilization effect on the cognate and near-cognate tRNAs (20).

After entering the ribosome, aa-tRNA must bent about 30° to interact with the mRNA codon and EF-Tu simultaneously, thereby adopting the A/T state (49). When the ribosome binds a cognate tRNA, it strongly stabilizes the mRNA-tRNA interaction, enabling the tRNA to overcome bending potential and maintain the bent state long enough for spontaneous elbow rotation and prepare for translocation (15). In contrast, near-/non-cognate tRNAs receive weaker stabilization, which makes it difficult for the tRNA to bend for a long time and dissociate quickly. This explains why all currently resolved near-cognate structures that successfully pass ribosomal proofreading employ base tautomerization to acquire enhanced stabilization.

Beyond the differences in mRNA-tRNA stability, does A1493 exhibit different responses to cognate and near-cognate tRNAs? To investigate this, we calculated the flipping free energy curves of A1493 in four systems (U-A, U-U, U-C, U-G), as shown in Figure 5D. The results revealed similar free energy curves for A1493 flipping in different systems, with angles distributed around 130°, indicating that A1493 flipping lacks specificity. In contrast, previous molecular dynamics simulations by Zeng et al. revealed a 3 kcal/mol difference in the flipping of A1492/3 between cognate and near-cognate tRNA binding (27). This inconsistency may arise from differences in modeling systems. The earlier simulations were based on the closed conformation of 30S subunit with tRNA in the A/A state, whereas our study models the recognition intermediate with open 30S subunit and tRNA in the A/T state (18). A significant difference between the two structures is that before 30S subunit closure, such as PDB: 5UYL, 5UYK (18), 5WFK (50) and 4V6G (51), A1913 from the large subunit inserted into helix 44 and A1493 flips out from helix 44 (Figure 7A). Conversely, after 30S subunit closure, A1913 move away from helix 44 and establishes interaction with the 37th nucleotide of the tRNA (Figure 7B). Among them, the crystal structure 4V6G represents a ribosomal complex with solely a P-site tRNA, lacking an A-site tRNA; whereas in 5UYK, the A-site tRNA has entered the ribosome but has not yet formed base-pairing with the A-site codon. These observations suggest that the insertion of A1913 and the flipping out of A1493 may be intrinsic properties, independent of whether the A-site tRNA exists and whether it is cognate tRNA.

Therefore, simulations based on the structure where the 30S subunit is closed and A1913 has already detached from helix 44 cannot at least represent the entire biological process of tRNA recognition. The previously calculated free energy difference associated with the flipping of A1493 may be eliminated by the insertion of A1913. Compared to A1492, A1493 first interacts with the codon-anticodon helix, and at least at this step, the structural changes of

A1493 are equivalent across different tRNAs according our results. Consequently, it is impossible to distinguish cognate tRNAs from near-cognate tRNAs based solely on A1493.

However, it is noteworthy that certain crystal structures, such as 5KPS and 5KPV (52), which represent states before and after A-site tRNA binding, respectively, depict A1493 in a flipped-out conformation even in the absence of A1913 insertion into helix 44. This even means that at some point the extrahelical flipping of A1493 may be independent of A1913 and A site tRNA. To comprehensively elucidate the microscopic pathways underlying ribosomal recognition of different tRNAs, future simulations must simultaneously incorporate the conformational dynamics of A1913 from the large subunit and A1493/A1492 from the small subunit. This approach should explore the processes of A1913 insertion into and moving away from helix 44, as well as the flip of A1493 and A1492 from helix 44 toward the codon-anticodon helix. These processes may encompass multiple pathways, each characterized by distinct energy barriers or differential specificity responses to various tRNAs.

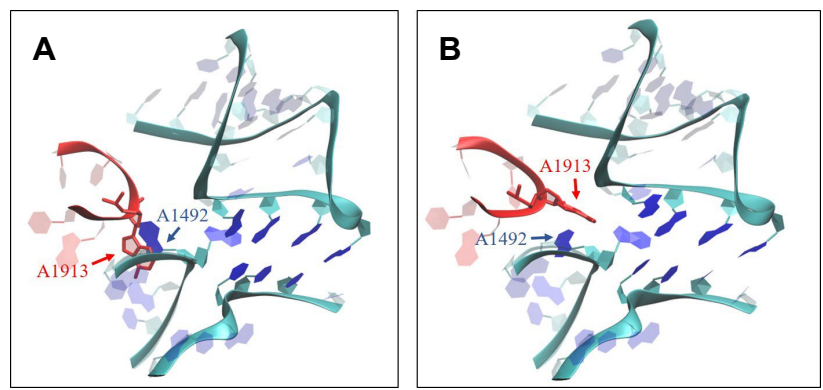

**Figure 7.** Two structures of A1913 (shown in red). (A) A1913 inserted into helix 44 and stack with A1492. (B) A1913 detached from helix 44 and interact with the sugar of 37th nucleotide of tRNA.

**Ψ enhances mRNA-tRNA pairing by pre-organizes A-form of mRNA**

In the experiment conducted by Koutmou et al., it was observed that a single substitution of uridine U with pseudouridine (Ψ) in the codon UUU results in amino acid substitutions (33). For instance, the ΨUU codon, where the first base U was replaced by Ψ, was observed to incorporate Val, Leu, and Ile; while the UUΨ codon, with the third base U replaced by Ψ, led to the incorporation of Val and Leu. Although their experiment was conducted under simulated starvation conditions, the solution contained both cognate tRNAs and near-cognate tRNAs simultaneously. The amino acid substitutions observed under these conditions cannot be explained by base tautomerism, because the extremely low probability of tautomerism is insufficient to account for the high frequency of near-cognate tRNA binding. Consequently, we simulated a ribosomal system in which the first base of the codon is substituted with Ψ, forming the ΨUU codon, bound to cognate or near-cognate tRNAs.

The free energie curves of Ψ paired with A, U, C, and G are presented in Figure 8A, represented by blue, orange, red, and green solid lines, respectively. The color-shaded areas indicate the difference in opening free energy compared to the corresponding unmodified systems. Computational analysis reveals that, despite the additional hydrogen bond donor in Ψ not directly participating in pairing with tRNA, the stability of all systems is significantly enhanced following the modification of U to Ψ. Notably, the opening free energies of Ψ-U and

Ψ-C pairs exhibit stability comparable to that of the unmodified U-A base pair. This suggests that the stability enhancement induced by Ψ can effectively compensate for the reduced stability in mismatched base pair. This may lead to a deceptive effect on the ribosome, whereby it misidentifies the modified mismatched base pairs as cognate U-A base pairs, accepting near-cognate tRNAs and resulting in amino acid substitutions. Regarding the flipping angle of A1493, the free energy profiles remain consistent with those observed prior to modification (Figure S4).

Previous studies on Ψ have indicated that its contribution to system stability varies depending on the contextual sequence and its position, generally enhancing stability in helical regions while destabilizing loop (53–57). When located in helices with ΨU/AA sequences, Ψ stabilizes base stacking energy by approximately 1 kcal/mol (55). However, this modest stabilization cannot fully account for the observed ribosomal stability enhancement. Notably, if the stabilizing effect of Ψ were solely attributed to enhanced base stacking interactions, its presence in the middle of the codon-anticodon helix (the second position) should provide the greatest stabilization. However, the probability of amino acid substitution in the UΨU codon is the lowest, suggesting that Ψ must additionally enhance codon-anticodon stability through alternative mechanisms. We found that the extra hydrogen bond donor N1 in Ψ base forms an indirect interaction with the phosphate group between Ψ4 and G3 via a water bridge, as illustrated in Figure 8B. This phosphate group is located at the P/A kink, and this interaction enables Ψ4 to resist the inherent flipping tendency imposed by the P/A kink, thereby enhancing the stability of the first base pair. The free energy curve of the Ψ-A base pair in the absence of the ribosome is shown in the gray dashed line in Figure 3A. In contrast to the unmodified U4 base (purple dashed line), which transitions into the open state at 6.7 Å, the modified Ψ base maintains a constrained planar conformation without rotation until reaching approximately 8 Å, where it finally adopts a more freely moving open state. In other words, Ψ pre-organizes the A-form helical structure of mRNA through water bridge, thereby enhancing the pairing stability between mRNA and tRNA.

Beyond water-mediated bridging, the additional N1 hydrogen bond donor in Ψ may engage in direct interactions with the hypermodified t6A side chain at tRNA position 37 in the Ψ-U paired Ile-tRNA system, conferring extra stability to the first base pair. For the Ψ-C paired Val-tRNA system, we observed displacement of the tRNA's C36 base rather than the Ψ4 base. This observation arises because the unmodified U-C pair represents the least stable system; Ψ-induced stabilization shifts the dynamics, resulting in the movement of tRNA's C36 instead. Therefore, to enable direct comparison with the unmodified U-C system, the free energy of the Ψ-C pair was calculated with C36 flipping constrained. While Ψ-G pairing exhibited relatively modest stabilization compared to other mismatches, it still induced significant amino acid misincorporation, suggesting Ψ modifications may influence ribosomal processes beyond the decoding center. For example, experimental evidence has demonstrated that Ψ modification induces disordering of the tRNA's 5'CCA end—a phenomenon that our current simulations cannot account for, indicating potential limitations in the computational model or additional mechanistic factors at play.

In the Ψ-A paired Phe-tRNA system, two distinct structures have been experimentally resolved, differing primarily in the glycosidic angle of A1493 (33, 37). In crystal structure 6UO1 and 4V9I, A1493 adopts anti conformation, consistent with the majority of resolved structures. Conversely, in crystal structure 4JV5, A1493 exhibits an unusual syn conformation, which is interpreted as that in the Ψ-A pairing, the syn state of A1493 may has a stronger A-minor interaction than anti state (37). To explore this hypothesis, we constructed two systems based on the crystal structures 4JV5 and 4V9I (Figure S5) and calculated the flipping free energy curves of A1493 in the syn and anti state, respectively (Figure S6). Our findings reveal that the flipping free energy barrier is slightly lower in the syn state than in the anti state, suggesting that the syn conformation does not favor A-minor interactions. We propose that the syn conformation observed in 4JV5 may arise due to the retention of only the 30S small ribosomal subunit in the crystal structure, lacking the critical contribution of A1913 from the 23S large subunit. During its insertion into helix 44, A1913 interacts with A1492 and A1493, potentially restricting excessive rotation of the A1493 glycosidic bond during flipping and maintaining it in the anti conformation. In addition, we observed that when A1913 detaches from helix 44, a stacking structure with A1493 may form, which indirectly supports our hypothesis. This observation once again highlights the critical role of A1913 in the large ribosomal subunit, underscoring the necessity of considering A1913's behavior in future investigations of the complete process of tRNA recognition by the ribosome.

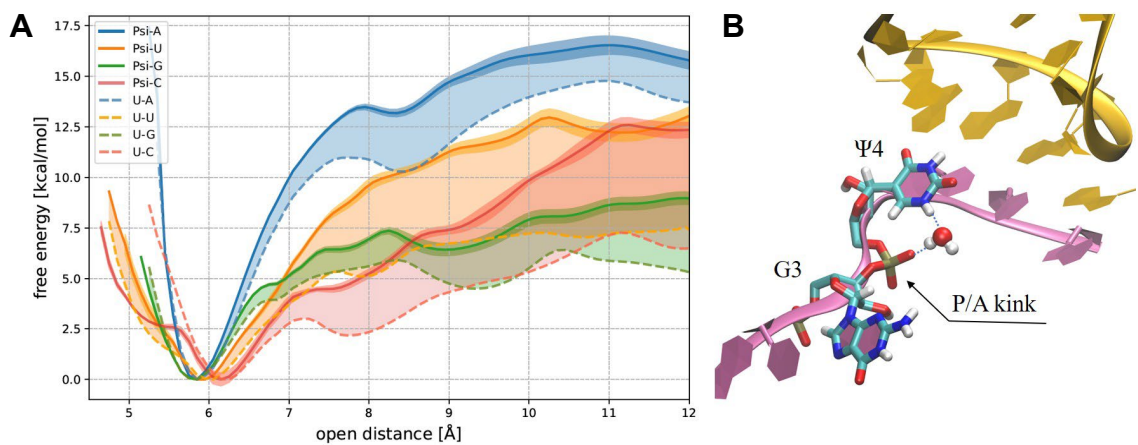

**Figure 8.** (A) The free energie curves of Ψ paired with A, U, C, and G. The color-shaded areas indicate the difference compared to the corresponding unmodified systems. (B) The water molecule bridging Ψ and phosphate group. The pink backbone is mRNA, and the gold backbone is tRNA.

**Stability changes of the codon-anticodon helix during translocation**

In the preceding sections, we elucidated the mechanism by which A1493 stabilizes the codon-anticodon helix at the ribosomal A-site. However, as the ribosome translocates to the P-site, the codon-anticodon helix progressively moves away from the critical A-site nucleotides A1492, A1493, and G530. This movement prompts the question: Does this displacement lead to destabilization of the mRNA-tRNA interaction during translocation and at the P-site? Our calculations show that the translocation process is indeed accompanied by instability, but translocation to the P site will be stabilized again.

Figure 9A shows the partial ribosome structure around the mRNA. The pink part is the mRNA and tRNA at the A site before translocation (PDB: 7SSD), and the yellow part is the mRNA and tRNA that have been translocated to the P site in the post-translocation stage (PDB: 7SSW) (15). Due to the relative rotation between the head and body of the small subunit during translocation, the blue ribosomal structure in the figure is the overlapped region of both states, excluding distant areas that undergo substantial conformational changes. We retained ribosomal components within 20 Å of P-site codon's first base pair in the post-translocation state and calculated the opening free energy, represented by the yellow line in Figure 9B. This energy barrier is nearly identical to that observed at the A-site before translocation (pink line), suggesting that the ribosome stabilizes this base pair at the P-site through a mechanism distinct from the A1493-mediated stabilization at the A-site.

The close view near the P-site codon is shown in the inset of Figure 9A. The rRNA beneath the mRNA is helix 44, which forms a local bulge at U1498. Here, U1498 flips outward from helix 44 and stacks onto U1497. Furthermore, the phosphate group of the U1498 forms a hydrogen bond with A790 from another ribosomal helix, enhancing the stability of this bulge structure. U1498 is close to the backbone of the first and second nucleotides of the P site codon, and forms extensive interactions such as hydrogen bonds. This significantly constrains the conformational of the mRNA, preventing the first codon base from moving away from the tRNA. Additionally, in contrast to the pronounced twist at the P/A kink, the twist at the E/P junction is relatively mild, providing further stabilization. During the separation of U4 from A36 at this site, U4 moves within a distance of 9.5 Å. However, once the mRNA backbone is restricted by U1498, U4 no longer moves, but A36 of tRNA continues to flip out, which is not observed in the A site. Since our analysis focuses solely on the opening free energy of the first base pair, and given that U1498 imposes spatial constraints on at least the first two nucleotides, it can be expected that the overall mRNA-tRNA interaction at the P-site is substantially more stable than at the A-site.

Due to the absence of crystal structural in the middle of the translocation path between the A site and the P site, we simulated the initial displacement by moving the mRNA-tRNA duplex along the A-to-P-site axis by the distance of one base stack within the ribosome (gray structure in Figure 9C), based on the A-site structure. We assumed that the ribosome has not yet changed during the initial movement of this translocation. In fact, structural rearrangements during translocation occur in regions distant from the mRNA, with minimal impact on the proximal mRNA vicinity, as evidenced in Figure S7. During this short displacement (approximately 4 Å), the first base pair of the codon-anticodon helix disengages from A1493 and has not yet entered the stabilizing range of U1498 at the P-site, without causing steric clashes between the mRNA or tRNA and the ribosome. The opening free energy of this base pair, shown by the gray line in Figure 9B, exhibits a significant reduction in the energy barrier, confirming that destabilization of the codon-anticodon helix occurs during translocation. At this stage, without the stabilizing effect of A1493, the first codon base becomes more prone to disrupting its pairing with the tRNA due to P/A kink. The vacated tRNA base may then pair with the subsequent codon—for example, the unpaired hydrogen atom of the NH2 group of tRNA's A36 could form a bifurcated hydrogen bond with the adjacent codon U5. This process propagates sequentially across the three anticodon bases until the base at position 34 disrupts its pairing with the third codon base and pairs with the base immediately following the codon (stacked in an A-form configuration on the codon helix), potentially resulting in a +1 frameshift. The propagation of A1493-mediated stabilization to the codon was also implied by the study of codon degeneracy by Ye et al. (66). Alternatively, the pairing between the preceding P site 3$^{rd}$ codon base, acting as a less stable wobble pair, may be disrupted, leading to pairing with the A-site tRNA and causing a -1 frameshift. Whether the reading frame shifts forward or backward, both outcomes depend on the destabilization of the codon-anticodon interaction. Our quantitative analysis provides a mechanistic basis for the experimentally observed disruptions in codon reading frames during translocation (58, 59). Notably, our calculations exclude elongation factor EF-G, as predicting structural changes in EF-G's flexible domain IV during translocation remains challenging. Yet the observed destabilization during translocation inversely highlights EF-G's critical role in maintaining codon-anticodon helix stability throughout this process.

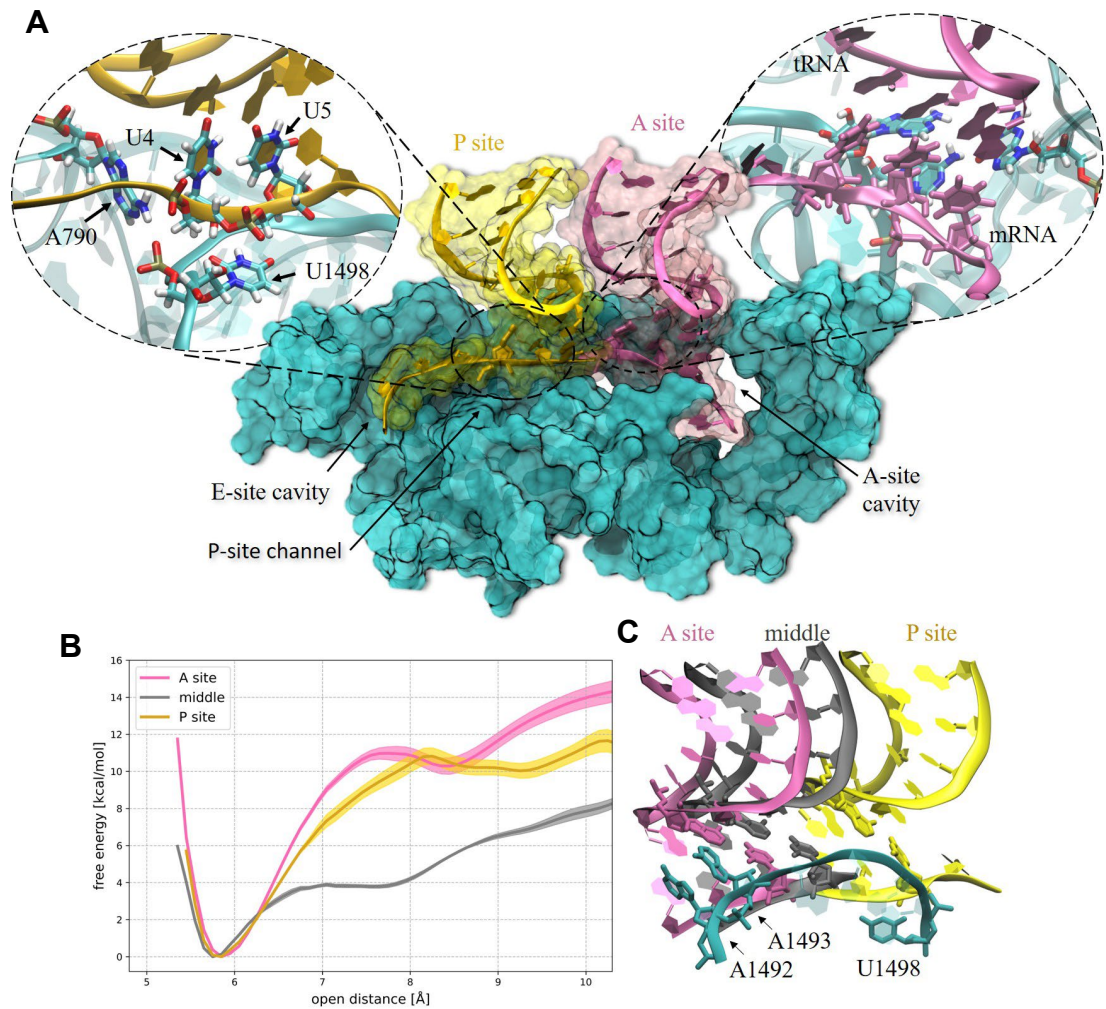

**Figure 9.** (A) Part of the ribosome near the mRNA. The blue part is rRNA; the pink part is the mRNA and tRNA at the A site; the yellow part is the mRNA and tRNA that have been translocated to the P site in the post-translocation stage; and the local magnified images of the two sites. (B-C) are the base pair opening free energy curves and structural changes in the A site (pink), P site (yellow), and during translocation (gray). The blue structure in Figure C is part of helix 44, and the nucleotides shown from left to right are A1492, A1493, and U1498.

## SUMMARY AND DISCUSSION

The mechanism of tRNA selection by the ribosome has been investigated for over half a century,

yet no definitive consensus has emerged. There are currently two models, each supported by different evidence. The first model proposes that conserved ribosomal nucleotides (G530, A1492, and A1493) stabilize interactions with the cognate codon-anticodon helix, triggering domain closure, GTP hydrolysis of EF-Tu, and tRNA accommodation into the 50S A-site. Conversely, these nucleotides fail to stabilize near-cognate codon-anticodon helices, preventing domain closure and resulting in tRNA dissociation. This hypothesis is supported by some crystal structures and computational studies (2, 21, 22, 27). The second model proposes that the ribosome interacts with the near-cognate tRNA in the same way as the cognate tRNA, which also triggers the same domain closure. The rigid decoding center forces non-canonical base pairs to adopt a Watson-Crick-like geometry, but the energetically costly base tautomerization leads to unfavorable overall energy, prompting tRNA dissociation. This view is also supported by some crystal structures and computational studies (24–26, 60, 61).

In this study, we performed molecular simulations using ribosomal recognition intermediates as structural templates to model a series of cognate and near-cognate tRNA-ribosome complexes and quantifying interactions among the A1493, mRNA, and tRNA. Our results provide support for some arguments of both models, suggesting that the two models are complementary.

First, our simulated structures support model one. Molecular dynamics simulations reveal that the U-G mismatch at the first position of the codon-anticodon helix adopts a wobble structure under the constraints of covalent linkage between A- and P-site codons and torsional restrictions from P/A kink. This finding contradicts the second model's claim that "Because it (2) does not have the natural restraint coming from being covalently bound to the P codon, this first nucleotide in the A codon has the freedom to move so it can form a wobble U-G pair" (25). We propose that the crystal structure of the second model represents a very low probability of successfully evading ribosome proofread under physiological conditions, while the crystal structure of the first model represents most of the structure before being rejected by the ribosome and dissociated. Especially for the U-G wobble base pair, there is no steric conflict with A1493 or A1492 after their extrahelical flip from helix 44 (unlike G-U, which will have steric conflict with A1493 if it forms a wobble structure), and it only has one fewer hydrogen bond with A1493 than the cognate base pair. Whether this can "force" the first position of U-G to undergo tautomerization is questionable, because studies have shown that the keto-enol transitions require an energy cost up to 10 kcal/mol (65). Classical force field studies suggest that the Watson-Crick U-G at the first position is unstable (27, 64), with one study noting stability at the middle position (64). Semi-empirical QM (PM7) analysis indicates that A1492 and A1493 in the closed state slightly favor a Watson-Crick over a wobble configuration for U-G at the middle position (63). This finding confirms the ribosome's active role in enforcing Watson-Crick structure at second position of codon. However, given the more complex electrostatic environment at the middle position (influenced by A1492, G530, and dual stacking effects), the role of A1493 in driving tautomerization at the first position remains uncertain and merits further exploration.

Second, our calculations reveal significant stability differences between cognate and near-cognate systems, with A1493 playing a critical role in stabilizing the codon-anticodon helix and amplifying free energy disparities. These findings support Model one's proposal that the ribosome discriminates tRNAs via structural-dependence of codon-anticodon stability.

Comparative studies with ribosome-free control systems demonstrated that A1493 stabilizes the system entropically by constraining the conformational numbers of the first codon nucleotide's open state. Further investigations showed reduced stabilization when A1493 is in the "off" state, its stabilizing effect is weakened because its sugar makes room for the active space of U4 in the open state, making it easier for U4 to enter the open state driven by entropy increase. This indicates that the maximum stabilization effect of the ribosome on tRNA depends on the complete extrahelical flipping of A1493.

Third, the free energy of A1493 flipping reveal nearly identical angles and energy profiles for cognate and near-cognate systems, indicating a lack of specificity. This does not support the view of the first model that A1493 has selective specificity and the previous simulation results (27). We attribute this to structural modeling differences: multiple crystal structures (18, 53) show that A1913 from the 50S large subunit inserted into helix 44 and A1493 flips out from helix 44. This A1913 insertion may eliminates free energy differences in A1493 flipping, supporting the second model's view that the ribosome cannot distinguish tRNAs based on A1493 flipping angles.

Fourth, from the coupling of the A1493 flip angle and the first base pair distance, we found that as A1493 moves away from the codon-anticodon helix, the mRNA and tRNA backbones tend to approach each other. This shows that when A1493 fully extrahelical flips, it can support the tRNA backbone like a "wedge", preventing the mRNA and tRNA backbones from getting too close, and providing rigid properties for the decoding center, which supports the second model's view that the decoding center has a rigid structure.

Fifth, we pointed out that the support of A1493 for the tRNA backbone comes from the vdW interaction between A1493 and the sugar at the 37th position of tRNA. Through control simulations that eliminate this interaction, we established its essential role in facilitating the correct flipping of A1493. This illustrates the structural role of the tRNA 37th site and expands the understanding of this site—previously thought to function solely in providing additional stabilization to the first base pair. The mRNA/tRNA nucleotide at the first position and the 37th position nucleotide of tRNA together constitute the space to accommodate A1493, becoming the structure basic to steric complementarity. This is consistent with the view that steric complementarity is important for the decoding center pointed out by two models and several experiments (28, 29).

In summary, we propose that the two models are complementary: the first model elucidates the mechanism when the ribosome accurately identifies near-cognate tRNAs, while the second model describes scenarios where near-cognate tRNAs escape proofreading. Additionally, the role of A1913 in the 50S large subunit prior to small subunit closure should be incorporated into the first model, as it influences A1493 flipping specificity.

In addition, we explored the effect of $\Psi$ modification on ribosomal recognition. When the first codon base is modified from U to $\Psi$, the additional N1 hydrogen bond donor of $\Psi$ bridges the P/A kink via a water molecule, stabilizing the base's orientation and mitigating the P/A kink's destabilizing influence. In other words, $\Psi$ enhances mRNA-tRNA pairing by pre-organizes A-form of mRNA. This even increased the stability of $\Psi$-C and $\Psi$-U mismatches to the level of U-A. This at least partially explains why amino acid substitutions are more likely to occur after modification.

Lastly, we calculated the stability changes of the first base pair during the translocation.

Our findings indicate that the codon-anticodon interaction undergoes destabilization during translocation. However, as the codon approaches the P-site, the ribosome restores stability by imposing constraints via U1498 on the first and second positions of the P-site codon. From the above analysis of the molecular mechanism back to the biological role of the ribosome, the codon at the A site needs to recognize and stably bind to tRNA, which requires the codon to be exposed to the solution and flexible like a probe to expand the capture range. Consequently, the ribosome employs a stabilization strategy that minimizes interference with codon flexibility by positioning the codon within a cavity (as depicted in Figure 9). Rather than exerting excessive constraints on the mRNA backbone, the ribosome spatially restricts the potential flipping of the first codon base. This approach stabilizes what would otherwise be the most unstable base due to the influence of the P/A junction. In contrast, at the P-site—where tRNA pairing and dissociation do not occur—the mRNA is confined to a narrow channel. This significantly limits the flexibility of the mRNA backbone, thereby enhancing stability. At the E-site, the ribosome forms a cavity to facilitate tRNA dissociation.

In this investigation of the ribosomal decoding mechanism, while our focus has been on the flipping of A1493 and the stability of the first-position base pair, the role of A1913 in the large subunit has emerged as a critical factor during analysis. Across various structures prior to tRNA binding at the A-site, A1493 was observed partially embedded within helix 44 and partially flipped out. In structures where A1493 was flipped out, A1913 partially inserted into helix 44 and partially not. These diverse configurations persisted even after tRNA binds to the A-site. These observations raise a fundamental question: Does A1493 flip out prior to A1913 insertion, or does A1913 insertion actively displace A1493? Alternatively, might both pathways coexist? The answer to this question will help determine whether A1493 flipping exhibits tRNA-specificity. To comprehensively elucidate the microscopic pathways underlying the ribosome's recognition of different tRNAs, it is essential to integrate the conformational changes of A1913, alongside those of A1493 and A1492 into simulation models. This requires simultaneous investigation A1913's insertion into and detachment from helix 44, as well as A1493/A1492 transitions between intra-helix conformations and extrahelical flipping toward the codon-anticodon helix.

Moreover, current simulation methods are constrained by several limitations, leaving numerous phenomena unexplained. For instance, simulations have been restricted to a small region of the ribosome, preventing analysis of the competition between the overall elastic potential energy of the bent tRNA and its complete interaction with mRNA. Additionally, the amplification and transmission of differences in the decoding center through the rod-like structure of tRNA, as well as the process of domain closure, remain unobservable. Resolving these challenges will require enhancements in computational speed and reductions in resource demands, or the development of more precise approximation techniques, such as coarse-graining or simplification of complex interactions (67). In the ongoing debate regarding the decoding center's mechanism for tRNA recognition, high-precision ab initio molecular dynamics simulations may be essential to determine whether base tautomerization occurs prior to or during decoding, its ease and frequency, and whether conserved nucleotides in the decoding center facilitate or suppress this process. The impact of RNA modifications on this mechanism also warrants investigation. Experimentally, capturing the structures of near/non-cognate tRNAs at the moment of dissociation from the decoding center may necessitate engineered designs of tRNA or the ribosome. Such designs could introduce enhanced interactions between tRNA and the ribosome outside the decoding center to prevent the dissociation of near/non-cognate tRNAs. A precise understanding of this decoding mechanism holds particular importance for the design and optimization of RNA-based therapies. Such insight could mitigate mistranslation events, which may compromise therapeutic efficacy or elevate toxicity.

## ACKNOWLEDGMENTS

This work was partly supported by the National Natural Science Foundation of China under Grant No. 11574234. The numerical calculations in this work were performed on the supercomputing system in the Super Computing Center of Wuhan University.

## Data Availability Statement

The data that support the findings of this study are available from the corresponding author upon reasonable request.